\documentclass[journal,twoside,final]{IEEEtran}
\usepackage{graphicx}
\usepackage{amsmath}
\usepackage{latexsym}
\usepackage{graphicx}
\usepackage{bm}
\usepackage{amssymb}
\usepackage{array}
\usepackage{setspace}
\usepackage{fancyhdr}
\usepackage{amsmath, amssymb, epsfig}
\usepackage{float}
\usepackage{subcaption}
\usepackage{balance}
\usepackage{color}

\usepackage{csquotes}
\MakeOuterQuote{"}

\makeatletter
\newcommand{\vast}{\bBigg@{3.2}}
\newcommand{\Vast}{\bBigg@{4.2}}
\makeatother
\newtheorem{theorem}{Theorem}

\def\proof{\noindent\hspace{2em}{\itshape Proof: }}
\def\endproof{\hspace*{\fill}~$\blacksquare$\par\endtrivlist\unskip}

\begin{document}
\title{\linespread{1} Interference Impact on DF Relay Networks with RIS-Assisted Source and Relays}

\author{Anas~M.~Salhab\authorrefmark{1},~\IEEEmembership{Senior~Member,~IEEE} and Liang Yang\authorrefmark{2}\\
%\vspace{-0.0cm}
\thanks{\authorrefmark{1} Anas M. Salhab is with the Department
of Electrical Engineering, King Fahd University of Petroleum \&
Minerals, Dhahran 31261, Saudi Arabia (e-mail: salhab@kfupm.edu.sa).}

\thanks{\authorrefmark{2} Liang Yang is with the College of Computer Science and Electronic
Engineering, Hunan University, Changsha 410082, China, and also with the
State Key Laboratory of Integrated Services Networks, Xidian University,
Xi’an 710126, China (e-mail: liangy@hnu.edu.cn).}}
%\vspace{-0.6cm}}

%\thanks{Manuscript received April 19, 2005; revised January 11, 2007.}

\maketitle{}

\begin{abstract}
%\boldmath
In this letter, we consider the scenario of decode-and-forward relay network with reconfigurable intelligent surface (RIS)-assisted source and relays in the presence of interference. We derive approximate closed-form expression for the system outage probability assuming Rayleigh fading channels and opportunistic relaying scheme. In addition, we study the system behavior at the high signal-to-noise ratio (SNR) regime, where the diversity order and coding gain are obtained and analyzed. The results show that the system can achieve a diversity order of $G_{d}=\mathrm{min}(N_{1},N_{2})K$, where $N_{1}$ and $N_{2}$ are the numbers of reflecting elements at the source and relays, respectively, and $K$ is the number of relays. In addition, findings illustrate that for the same diversity order, utilizing one relay with multiple reflecting elements gives better performance than utilizing multiple relays each with a single reflecting element. Furthermore, findings illustrate that the interference at the destination is more severe on the system performance than the interference at the relays. Therefore, under the same interference powers and for a fixed number of relays $K$, results show that the case where the first hop is dominating the performance $N_{1}<N_{2}$ gives better results than the case where $N_{2}<N_{1}$ in terms of coding gain.

\end{abstract}
\textbf{\small \emph{Index Terms\textemdash}Reconfigurable intelligent surface, decode-and-forward relay, Rayleigh fading, co-channel interference.}


% \fi
\IEEEpeerreviewmaketitle

%%%%%%%%%%%%%%%%%%%%%%%%%%%%%%%%%%%%%%%%%%%%%%%%%%%%%%%%%%%%%%%%%%%%
%%%%%%%%%%%%%%%%%%%%% Section 1: Introduction %%%%%%%%%%%%%%%%%%%%%%
%%%%%%%%%%%%%%%%%%%%%%%%%%%%%%%%%%%%%%%%%%%%%%%%%%%%%%%%%%%%%%%%%%%%

\section{Introduction}
Owing to their excellent features, the reconfigurable intelligent surfaces (RISs) have recently attracted a noticeable attention as a promising technique for future wireless communication networks. 
An RIS is an artificial surface, made of electromagnetic
material, that is capable of customizing the propagation of
the radio waves impinging upon it \cite{Renzo1}. 
It has been proposed as a new low-cost and less complicated solution to realize wireless
communication with high energy and spectrum efficiencies \cite{Basar1}.

In \cite{Wu}, it was shown that RIS outperforms
conventional massive multiple-input multiple-output
systems and multi-antenna
amplify-and-forward relaying networks with smaller
number of antennas, while reducing the system complexity and
cost. The performances of relay-assisted and RIS-assisted
wireless networks from coverage, probability of signal-to-noise ratio (SNR) gain, and delay outage rate aspects have been compared in \cite{Yang2}. In \cite{Yang4}, the authors utilized RIS to improve 
the quality of a source signal that is sent to destination through an unmanned aerial vehicle.

The authors in \cite{Yang6} and \cite{Boulogeorgos} derived accurate approximations for channel distributions of RIS-assisted networks assuming Raleigh fading. Recently, some works on RIS-assisted networks over Nakagami-$m$ fading channels started to appear in literature \cite{Ferreira}, \cite{Anas1}. Most recently, limited number of papers have considered the interference phenomenon in a RIS context \cite{Li}, \cite{Hou}. All these works on interference effect considered the scenario, where an RIS is used as a replacement for relay.

There exists another important scenario where the RIS could be used as a part of the transmitter itself \cite{Basar1}, \cite{Yang6}. In such scenario, the transmitter includes an RF signal generator and an RIS, where the RIS is used as a transmitter along with the RF signal generator \cite{Basar1}. Recently, this idea was validated with the aid of a testbed platform \cite{Tang}. Combining RIS with relays is expected to achieve very high diversity order. In addition, addressing the impact of interference on the behavior of such scenario is of a noticeable importance.

In this letter, we consider the scenario, where the source and relays are internally utilizing RISs to enhance the quality of their transmitted signals in a decode-and-forward (DF) relaying network with interference. We derive accurate closed-form approximation for the system outage probability assuming Rayleigh fading channels for both intended users and interferers. The derived results are valid for arbitrary number of reflecting elements. We also derive closed-form expression for the asymptotic outage probability at high SNR values, where the system diversity order and coding gain are provided. To the best of authors’ knowledge, the derived expressions are new and the achieved key findings on the impact of interference on the performance of RIS-assisted sources and relays network are being reported for the first time here.

%This paper is organized as follows. Section \ref{SM} presents the
%system model. The system performance is analyzed in Section
%\ref{PA}. After that, Section \ref{AOB} proposes the asymptotic
%outage behavior of the considered system. In Section \ref{NR},
%some simulation and numerical results are presented and discussed.
%Finally, conclusions are given in Section \ref{C}.

%%%%%%%%%%%%%%%%%%%%%%%%%%%%%%%%%%%%%%%%%%%%%%%%%%%%%%%%%%%%%%%%%%%%%%%%%%%
%%%%%%%%%%%%%%%%%%%%%%%%% Section 2: System Model %%%%%%%%%%%%%%%%%%%%%%%%%
%%%%%%%%%%%%%%%%%%%%%%%%%%%%%%%%%%%%%%%%%%%%%%%%%%%%%%%%%%%%%%%%%%%%%%%%%%%
\section{System and Channel Model}\label{SM}
Consider a dual-hop relay system with RIS-assisted source of $N_{1}$ reflecting elements, $K$ RIS-assisted DF relay nodes each of $N_{2}$ reflecting elements, one destination, and arbitrary number of interferers at both the relays and destination with opportunistic relaying. The entire communication takes place in two phases. In the first phase, the source {\sf S} transmits its signal to $K$ relays. In the second phase, only the best relay among all other relays who succeeded in decoding the source signal in the first phase is selected to forward it to destination {\sf D}. We assume that the signal at the $k^{\mathrm{th}}$ relay is corrupted by interfering signals from
$I_{k}$ co-channel interferers $\{x_{i}\}_{i=1}^{I_{k}}$.
%\begin{figure}[!htb]
%\centering
%\includegraphics[scale=0.48]{systema}
%\caption{\small A schematic for the dual-hop opportunistic DF
%relying system with CCI at both the relay and the
%destination.}\label{system}
%\end{figure}

With assuming that only the signals reflected by the RIS once are considered while those reflected two times and more are ignored, the received signal at the $k^{\mathrm{th}}$ relay can be expressed as
\begin{equation}\label{Eq.1}
y_{{\sf r}_{k}}= \sum_{i = 1}^{N_{1}} h_{{\sf s},k,i}e^{j\phi_{{\sf s},k,i}}x_{0}+\sum_{i_{k}=1}^{I_{k}}h_{ i_{k},k}^{I}x_{i_{k},k}^{I}+n_{{\sf
s},k},
\end{equation}
where $h_{{\sf s},k,i}=\alpha_{{\sf s},k,i}e^{-j\theta_{{\sf s},k,i}}$ is the channel coefficient between the $i^{\mathrm{th}}$ reflecting element at $\sf S$ and the $k^{\mathrm{th}}$ relay, where $\alpha_{{\sf s},k,i}$ is the amplitude and $\theta_{{\sf s},k,i}$ is the phase shift, $\phi_{{\sf s},k,i}$ is the adjustable phase induced by the $i^{\mathrm{th}}$ reflecting
element, $x_{0}$ is the transmitted symbol with $\mathbb{E}\{|{x_{0}}|^{2}\}=P_{0}$, $h_{i_{k},k}^{I}$ is the channel coefficient between the $i_{k}^{\mathrm{th}}$ interferer and $k^{\mathrm{th}}$ relay, $x_{i_{k},k}^{I}$ is the transmitted symbol from the $i_{k}^{\mathrm{th}}$ interferer with
$\mathbb{E}\{|{x_{i_{k},k}^{I}}|^{2}\}=P_{i_{k},k}^{I}$, $n_{{\sf s},k}\thicksim\mathcal{CN}(0, N_{0})$ is an additive white Gaussian noise (AWGN) of zero mean and power $N_{0}$, and $\mathbb{E}\{\cdot\}$ denotes the expectation operation. Let us define $h_{k,{\sf d},i}=\beta_{k,{\sf d},i}e^{-j\theta_{k,{\sf d},i}}$, where $\beta_{k,{\sf d},i}$ is the amplitude and $\theta_{k,{\sf d},i}$ is the phase shift, and ${h_{i_{d},\sf d}^{I}}$ as the channel coefficients between the $i^{\mathrm{th}}$ reflecting
element at the $k^{\mathrm{th}}$ relay and {\sf
D}, and the $i_{d}^{\mathrm{th}}$ interferer and {\sf D}, respectively. The amplitudes of the channels between {\sf S} and the $k^{\mathrm{th}}$ relay ${\alpha_{{\sf s},k,i}}, k=1...K$ and between the $k^{\mathrm{th}}$ relay and {\sf D} ${\beta_{k,{\sf d},i}}, k=1...K$ are assumed to be Rayleigh distributed with mean $\frac{\sqrt{\pi}}{2}$ and variance $\frac{(4-\pi)}{4}$. That is, their mean powers $\mathbb{E}\{|\alpha_{{\sf s},k,i}|^{2}\}=\mathbb{E}\{|\beta_{k,{\sf d},i}|^{2}\}=1$. In addition, the interferers channel coefficients are assumed to follow Rayleigh distribution. That is, the channel powers denoted by $|{h_{i_{k},k}^{I}}|^2$ and $|{h_{i_{d},\sf d}^{I}}|^2$ are
exponential distributed random variables (RVs) with parameters $\sigma_{I,i_{k},k}^{2}$ and
$\sigma_{I,i_{d},\sf d}^{2}$, respectively. Using (\ref{Eq.1}),
the signal-to-interference-plus-noise ratio (SINR) at the $k^{\mathrm{th}}$ relay can be written as
\begin{equation}
\gamma_{{\sf s},k}=\frac{P_{0}}{N_{0}}{\left(\sum_{i=1}^{N_{1}}|\alpha_{{\sf
s},k,i}|\right)^{2}}\Big/\left(\sum_{i_{k}=1}^{I_{k}}\frac{P_{i_{k},k}^{I}}{N_{0}}|{h_{
i_{k},k}^{I}}|^2+1\right).
\end{equation}

Let $B_{L}$ denote a decoding set defined by the set of relays who
successfully decoded the source message at the first phase. It is
defined as
\begin{align}\label{Eq.3}
B_{L}&\triangleq\left\{k\in \mathcal{S}_{r}:\gamma_{{\sf s},k}\geq
2^{2R}-1\right\},
\end{align}
where $\mathcal{S}_{r}$ is the set of all relays and $R$ denotes a
fixed spectral efficiency threshold.

In the second phase and after decoding the received signal, only the best relay in $B_{L}$ forwards the
re-encoded signal to the destination. The best
relay is the relay with the maximum
$\gamma_{l,\sf d}$, where $\gamma_{l,\sf d}$ is the SINR at the destination resulting from the $l^{\mathrm{th}}$ relay being the
relay, which forwarded the source information. It can be written as
\begin{equation}
\gamma_{l,\sf d}=\frac{P_{l}}{N_{0}}\left(\sum_{k=1}^{N_{2}}|{\beta_{l,{\sf
d},k}}|\right)^{2}\Big/\left(\sum_{i_{d}=1}^{I_{d}}\frac{P_{i_{d},\sf
d}^{I}}{N_{0}}\big|{h_{ i_{d},\sf d}^{I}}\big|^2+1\right),
\end{equation} where
$P_{l}$, $P_{i_{d},\sf d}^{I}$, and $N_{0}$ are the transmit power
of the $l^{\mathrm{th}}$ active relay, the transmit power of the $i_{d}^{\mathrm{th}}$ interferer, and the AWGN power at the
destination, respectively, and $I_{d}$ is the number of
interferers at the destination node. Since the denominator is
common to the SINRs from all relays belonging to $B_{L}$, the
best relay is the relay with the
maximum $\frac{P_{l}}{N_{0}}\left(\sum_{k=1}^{N_{2}}|{\beta_{l,{\sf
d},k}}|\right)^{2}$.

The end-to-end (e2e) SINR
at {\sf D} can be written as
\begin{equation}\label{dsinr}
\gamma_{\sf
d}=\frac{P_{b}}{N_{0}}\left(\sum_{k=1}^{N_{2}}|{\beta_{b,{\sf
d},k}}|\right)^{2}\bigg/\left(\sum_{i_{d}=1}^{I_{d}}\frac{P_{i_{d},\sf
d}^{I}}{N_{0}}\big|{h_{ i_{d},\sf d}^{I}}\big|^2+1\right),
\end{equation}
where the subscript $b$ is used to denote the best selected relay.
%%%%%%%%%%%%%%%%%%%%%%%%%%%%%%%%%%%%%%%%%%%%%%%%%%%%%%%%%%%%%%%%%%%%%%%%%%%%
%%%%%%%%%%%%%%%%%%%% Section 3: Performance Analysis %%%%%%%%%%%%%%%%%%%%%%%
%%%%%%%%%%%%%%%%%%%%%%%%%%%%%%%%%%%%%%%%%%%%%%%%%%%%%%%%%%%%%%%%%%%%%%%%%%%%
\section{Outage Performance Analysis}\label{PA}

\subsection{Preliminary Study}
The probability of the decoding set defined in (\ref{Eq.3}) can be
written as
\begin{align}\label{Eq.8}
\mathrm{P_{r}}\left[B_{L}\right]=\prod_{l\in
B_{L}}\mathrm{P_{r}}\left[\gamma_{{\sf s},l}\geq
u\right]\prod_{m\notin B_{L}}\mathrm{P_{r}}\left[\gamma_{{\sf
s},m}< u\right],
\end{align}
where $u=\left(2^{2R}-1\right)$ is the outage SNR threshold.
The outage probability of the system can be achieved by averaging
over all the possible decoding sets as follows \cite{JKim}
\begin{align}\label{Eq.9}
P_{\sf out}&\triangleq\mathrm{P_{r}}\left[\frac{1}{2}\
\mathrm{log}_{2}\left(1+\gamma_{\sf d}\right)<R\right]\nonumber\\
&=\sum_{L=0}^{K}\sum_{B_{L}}\mathrm{P_{r}}\left[\gamma_{\sf
d}<u|B_{L}\right]\mathrm{P_{r}}\left[B_{L}\right],
\end{align}
where the internal summation is taken over all of
${{K}\choose{L}}$ possible subsets of size $L$ from the set with
$K$ relays. In order to evaluate (\ref{Eq.9}), we need first to
derive $\mathrm{P_{r}}\left[\gamma_{\sf d}<u|B_{L}\right]$ and
$\mathrm{P_{r}}\left[B_{L}\right]$, which are presented in the
following section.

Throughout the analysis below, it assumed that $\rho=
P_{0}/N_{0}=P_{l}/N_{0}$ and $\rho_{I}=
P_{i_{k},k}^{I}/N_{0}=P_{i_{d},\sf d}^{I}/N_{0}$. The terms
$\rho_{I}|h_{i_{k},k}^{I}|^{2}$ and
$\rho_{I}|h_{i_{d},\sf d}^{I}|^{2}$ are exponential distributed
with parameters $\lambda_{i_{k},k}^{I}=1/\rho_{I}\sigma_{I,i_{k},k}^{2}$ and
$\lambda_{i_{d},\sf d}^{I}=1/\rho_{I}\sigma_{I,i_{d},\sf d}^{2}$. As the channels between {\sf S} and all relays and from relays to {\sf D} are assumed to have the same mean power, which equals 1, their average SNRs will be equal to $\rho=
P_{0}/N_{0}=P_{l}/N_{0}$ and their parameters will be equal to $\lambda_{{\sf s},k}=\lambda_{k,\sf d}=1/\rho, k=1,...,K$.

%%%%%%%%%%%%%%%%%%%%%%%%%%%%%%%%%%%%%%%%%%%%%%%%%%%%%%%%%%%%%%%%%%%%%%%
%%%%%%%%%%%%%%%%%%%% Subsection: i.n.d. Interferers %%%%%%%%%%%%%%%%%%%
%%%%%%%%%%%%%%%%%%%%%%%%%%%%%%%%%%%%%%%%%%%%%%%%%%%%%%%%%%%%%%%%%%%%%%%
\subsection{Outage Probability}
The approximate outage probability of the considered system is summarized in the
following key result.

\begin{theorem}\label{theorem:1}
The outage probability of RIS-assisted source and relays DF relay network with independent identically distributed (i.i.d.) second hop cumulative distribution functions (CDFs) ($\lambda_{1,{\sf d}}=\lambda_{2,{\sf d}}=...=\lambda_{{\sf r,d}}$) and i.i.d. interferers' powers at {\sf D} ($\lambda^{I}_{1,{\sf d}}=\lambda^{I}_{2,{\sf d}}=...=\lambda^{I}_{I_{d},{\sf d}}=\lambda^{I}_{{\sf d}}$) and at relays ($\lambda^{I}_{1,k}=\lambda^{I}_{2,k}=...=\lambda^{I}_{I_{k},k}=\lambda^{I}_{k}, k=1,...,K$) can be obtained in a closed-form expression by using \eqref{Eq.9}, after evaluating the
terms $\mathrm{P_{r}}\left[\gamma_{\sf d}<u|B_{L}\right]$ and
$\mathrm{P_{r}}\left[\gamma_{{\sf s},k}<u\right]$ as follows
\end{theorem}
\begin{align}\label{Eq.27az}
&\mathrm{P_{r}}\left[\gamma_{\sf
d}<u|B_{L}\right]= - \frac{\left(\lambda_{\sf
 d}^{I} \right)^{I_{d}}}{\left(I_{d} - 1 \right)!}e^{\lambda_{\sf r,d}^{I}} \left(-1 \right)^{I_{d}}\sum_{g=0}^{I_{d} - 1}\ \begin{pmatrix}
I_{d} - 1 \\ g
\end{pmatrix} (-1)^{g} \nonumber\\&\times\sum_{k=0}^{L} \begin{pmatrix}
L \\ k \end{pmatrix} (-1)^{k}  \sum_{j_{0}=0}^{N_{2}-1}...\sum_{j_{k}=0}^{N_{2}-1} \frac{u^{\sum_{n=0}^{k}j_{n}}}{(C/\lambda_{\sf r,d})^{\sum_{n=0}^{k}j_{n}}\prod_{n=0}^{k}j_{n}!}\nonumber\\
&\times\left( \lambda^{I}_{\sf d}+ u\lambda_{\sf r,d} k \right)^{-g - \sum_{n=0}^{k}j_{n} - 1} \Gamma\left(g+\sum_{n=0}^{k}j_{n}+1, \lambda^{I}_{\sf d}+ \frac{u\lambda_{\sf r,d} k}{C}\right),
\end{align}
\begin{align}\label{Eq.27bz}
&\mathrm{P_{r}}\left[\gamma_{{\sf
s},k}<u\right]\approx \frac{\left(\lambda_{k}^{I} \right)^{I_{k}}}{\left(I_{k} - 1 \right)!}e^{\lambda_{k}^{I}} \left(-1 \right)^{I_{k}}\sum_{g=0}^{I_{k} - 1}\ \begin{pmatrix}
I_{k} - 1 \\ g
\end{pmatrix} (-1)^{g} \nonumber\\
 &\times \vast(- (\lambda^{I}_{k})^{-g-1} \Gamma\left(g+1,\lambda^{I}_{k}\right)+ \sum_{i = 0}^{N_{1} - 1} \frac{u^{i}  (\lambda_{{\sf s},k})^{i}}{B^i i!\left(\lambda^{I}_{k} + \frac{u  \lambda_{{\sf s},k}}{B}\right)^{g+i+1}} \nonumber\\
 &\times\Gamma\left(g+i+1,\lambda^{I}_{k} + \frac{u \lambda_{{\sf s},k}}{B }\right)\vast),
\end{align}
where $\Gamma(.,.)$ is the upper incomplete gamma function \cite[Eq. (8.352.2)]{Grad.}.

\proof  See Appendix \ref{appendix:1}.\endproof

%Now, following the same procedure as in Appendix B, the
%CDF $\gamma_{{\sf s},k}$ can be evaluated in a closed-form
%expression as
%
%
%Having $\mathrm{P_{r}}\left[\gamma_{\sf d}<u|C_{L}\right]$ and
%(\ref{Eq.57}) being evaluated, the outage probability in
%(\ref{Eq.9}) can be obtained.
%%%%%%%%%%%%%%%%%%%%%%%%%%%%%%%%%%%%%%%%%%%%%%%%%%%%%%%%%%%%%%%%%%%%%%%%%%%%
%%%%%%%%%%%%%%%%%%%%% Section 4: Asymptotic Analysis %%%%%%%%%%%%%%%%%%%%%%%
%%%%%%%%%%%%%%%%%%%%%%%%%%%%%%%%%%%%%%%%%%%%%%%%%%%%%%%%%%%%%%%%%%%%%%%%%%%%
\section{Asymptotic Outage Behavior}\label{AOB}
Here, $P_{\sf out}$ can be written as $P_{\sf
out}\approx\left(G_{c}\rho\right)^{-G_{d}}$, where $G_{c}$ is the coding gain and $G_{d}$ is the diversity order \cite{Alouinib}.

\begin{theorem}\label{Theorem:2}
The asymptotic outage probability for RIS-assisted source and relays DF relay network with interference can be obtained in a closed-form expression by using \eqref{Eq.9}, after evaluating the
terms $\mathrm{P_{r}}\left[\gamma_{\sf d}<u|B_{L}\right]$,
$\mathrm{P_{r}}\left[\gamma_{{\sf s},k}<u\right]$ as follows
\begin{align}\label{Eq.28a}
&\mathrm{P_{r}}\left[\gamma_{\sf
d}<u|B_{L}\right]\approx - \frac{\left(\lambda_{\sf
 d}^{I} \right)^{I_{d}}e^{\lambda_{\sf d}^{I}}\left(-1 \right)^{I_{d}}}{\left(I_{d} - 1 \right)!C^{N_{2}L}(N_{2}!)^{L}} \sum_{g=0}^{I_{d} - 1}\ \begin{pmatrix}
I_{d} - 1 \\ g
\end{pmatrix}\nonumber\\
&\times (-1)^{g} (\lambda_{\sf r,d})^{N_{2}L} u^{N_{2}L} (\lambda_{\sf d}^{I})^{-g-N_{2}L-1} \Gamma\left(g+N_{2}L+1, \lambda^{I}_{\sf d}\right),
\end{align}
\begin{align}\label{Eq.28b}
&\mathrm{P_{r}}\left[\gamma_{{\sf
s},k}<u\right]\approx - \frac{\left(\lambda_{k}^{I} \right)^{I_{k}}e^{\lambda_{k}^{I}} \left(-1 \right)^{I_{k}}}{\left(I_{k} - 1 \right)!B^{N_{1}}N_{1}!}\sum_{g=0}^{I_{k} - 1}\ \begin{pmatrix}
I_{k} - 1 \\ g
\end{pmatrix} (-1)^{g} \nonumber\\&\times (\lambda_{{\sf s},k})^{N_{1}} u^{N_{1}} (\lambda_{k}^{I})^{-g-N_{1}-1} \Gamma\left(g+N_{1}+1, \lambda^{I}_{k}\right).
\end{align}
\end{theorem}
\proof See Appendix \ref{appendix:3}.\endproof 

Two extreme cases have been noticed here as follows.
\begin{itemize}
\item Case 1: $N_{1}<N_{2}$\\
Here, $P_{\sf out}$ is dominated by the first hop as follows
\begin{align}\label{Eq.28bss}
P_{\sf out}^{\infty}&\approx \vast(\vast[- \frac{\left(\lambda_{\sf r}^{I} \right)^{I_{r}}e^{\lambda_{\sf r}^{I}} \left(-1 \right)^{I_{r}}}{\left(I_{r} - 1 \right)!B^{N_{1}}N_{1}!}\sum_{g=0}^{I_{r} - 1}\ \begin{pmatrix}
I_{r} - 1 \\ g
\end{pmatrix} (-1)^{g}\nonumber\\
&\times u^{N_{1}} \frac{\Gamma\left(g+N_{1}+1, \lambda^{I}_{\sf r}\right)}{(\lambda_{\sf r}^{I})^{g+N_{1}+1}}\vast]^{-1/N_{1}}\rho\vast)^{-N_{1}K}.
\end{align}
Clearly, the diversity order of the system for this case is $G_{d}=N_{1}K$ and the coding gain $G_{c}$ is a function of several parameters, including $N_{1}$, $I_{r}$, $\lambda_{\sf r}^{I}$, and $u$.

\item Case 2: $N_{2}<N_{1}$\\
Here, $P_{\sf out}$ is dominated by the second hop as follows
\begin{align}\label{Eq.28ass}
P_{\sf out}^{\infty}&\approx \vast(\vast[ - \frac{\left(\lambda_{\sf
 d}^{I} \right)^{I_{d}}e^{\lambda_{\sf d}^{I}}\left(-1 \right)^{I_{d}}u^{N_{2}L}}{\left(I_{d} - 1 \right)!C^{N_{2}L}(N_{2}!)^{L}} \sum_{g=0}^{I_{d} - 1}\ \frac{\begin{pmatrix}
I_{d} - 1 \\ g
\end{pmatrix}}{(-1)^{-g}} \nonumber\\
&\times \frac{\Gamma\left(g+N_{2}L+1, \lambda^{I}_{\sf d}\right)}{(\lambda_{\sf d}^{I})^{g+N_{2}L+1}}\vast]^{-1/N_{2}L}\rho\vast)^{-N_{2}L}.
\end{align}
The second hop is dominant when the number of active relays is equal to number of all available relays, that is $L=K$. This means that the diversity order of this system for this case is equal to $G_{d}=N_{2}K$, whereas it is clear from \eqref{Eq.28ass} that the coding gain $G_{c}$ is a function of several parameters, including $N_{2}$, $I_{d}$, $\lambda_{\sf d}^{I}$, and $u$. 
\end{itemize}

Therefore, the diversity order of the system when $N_{1}$ is not equal to $N_{2}$ is $G_{d}=\mathrm{min}(N_{1},N_{2})K$. On the other hand, when $N_{1}=N_{2}=N$, the interference value at the relays and destination specifies, which hop dominates the system performance. The hop that suffers from higher interference dominates the system performance by determining its coding gain. When the interference at both hops is equal, we noticed that the second hop dominates the system performance through determining its coding gain as in Case 2 above. For the last two cases, the diversity order is $G_{d}=NK$.
%%%%%%%%%%%%%%%%%%%%%%%%%%%%%%%%%%%%%%%%%%%%%%%%%%%%%%%%%%%%%%%%%%%%%%%%%%%%%%%%%%%%%%
%%%%%%%%%%%%%%%%%%%%%%%%%%%% Section 5: Numerical Results %%%%%%%%%%%%%%%%%%%%%%%%%%%%
%%%%%%%%%%%%%%%%%%%%%%%%%%%%%%%%%%%%%%%%%%%%%%%%%%%%%%%%%%%%%%%%%%%%%%%%%%%%%%%%%%%%%%
\section{Numerical Results}\label{NR}
\begin{figure}[htb!]
\centering
\includegraphics[scale=0.34]{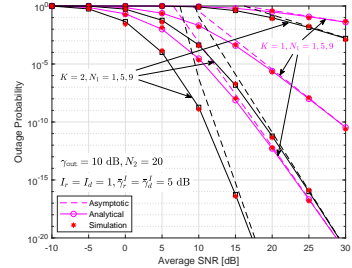}
\caption{$P_{\sf out}$ vs SNR for various values of $N_{1}$ and $K$.}\label{Pout_SNR_N1_K}
\end{figure}

We assume here that the parameters $\sigma_{k,\sf d}^{2}$, $\sigma_{I,i_{k},k}^{2}$, and
$\sigma_{I,i_{d},\sf d}^{2}$ are all equal to 1. In addition, we call the outage SNR threshold as $\gamma_{\sf out}=u$. A good matching between the analytical and asymptotic results and simulations is clear in Fig. \ref{Pout_SNR_N1_K}. In addition, the diversity order of the system $G_{d}$ is increasing when both $N_{1}$ and $K$ increase. As the system performance is dominated here by the first hop, $G_{d}=N_{1}K$ and this coincides with Case 1 in the asymptotic analysis section.

\begin{figure}[htb!]
\centering
\includegraphics[scale=0.34]{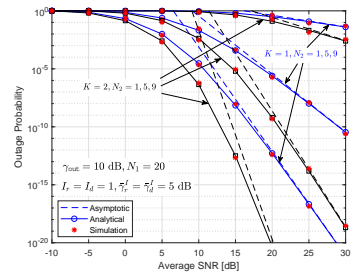}
\caption{$P_{\sf out}$ vs SNR for various values of $N_{2}$ and $K$.}\label{Pout_SNR_N2_K}
\end{figure}

Same insights can be drawn from Fig. \ref{Pout_SNR_N2_K}, but here $G_{d}=N_{2}K$ as the performance is dominated by the second hop. Results match Case 2 in the asymptotic analysis section. 
\restylefloat{figure}

\begin{figure}[htb!]
\centering
\includegraphics[scale=0.34]{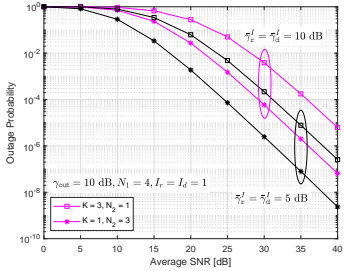}
\caption{$P_{\sf out}$ vs SNR for various values of $K$, $N_{2}$, $\bar{\gamma}_{\sf r}^{I}$, and $\bar{\gamma}_{\sf d}^{I}$.}\label{Pout_SNR_K_N2}
\end{figure}

In Fig. \ref{Pout_SNR_K_N2}, as $N_{2}$ is smaller than $N_{1}$, $G_{d}=3$. This figure informs us that the system with one relay node and multiple reflecting elements ($K=1,N_{2}=3$), gives better performance than the system with multiple relays each with one reflecting element ($K=3,N_{2}=1$).

\begin{figure}[htb!]
\centering
\includegraphics[scale=0.34]{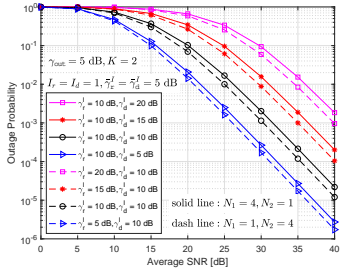}
\caption{$P_{\sf out}$ vs SNR for various values of $\bar{\gamma}_{\sf r}^{I}$ and $\bar{\gamma}_{\sf d}^{I}$.}\label{Pout_SNR_gIr_gId}
\end{figure}

Fig. \ref{Pout_SNR_gIr_gId} shows two main cases, (solid line: $N_{1}=4, N_{2}=1$) where the behavior is dominated by the second hop and (dash line: $N_{1}=1, N_{2}=4$) where the behavior is dominated by the first hop. For all curves here, $G_{d}=2$. For each main case, decreasing the interference power enhances $G_{c}$ and the system performance. In addition, this figure informs us that the interference at the destination is more impactful/severe on the system performance than the interference at the relays. This is clear as the case where ($\bar{\gamma}_{\sf r}^{I}=10\ \mathrm{dB},\bar{\gamma}_{\sf d}^{I}=20\ \mathrm{dB}$) gives worse results than the case where ($\bar{\gamma}_{\sf r}^{I}=20\ \mathrm{dB},\bar{\gamma}_{\sf d}^{I}=10\ \mathrm{dB}$).

\begin{figure}[htb!]
\centering
\includegraphics[scale=0.34]{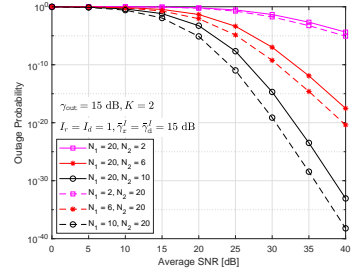}
\caption{$P_{\sf out}$ vs SNR for various values of $N_{1}$ and $N_{2}$.}\label{Pout_SNR_N1_N2}
\end{figure}

Fig. \ref{Pout_SNR_N1_N2} shows two cases, first hop dominating the system performance ($N_{1}<N_{2}$) and second hop dominating the system performance ($N_{2}<N_{1}$). For each case, increasing the number of reflecting elements $N_{1}$ or $N_{2}$ increases $G_{d}$ and enhances the performance. For the first case, $G_{d}=N_{1}K$ and for the second case, $G_{d}=N_{2}K$. Also, having more reflecting elements at the relays than the source gives better performance. This is because when the first hop is dominating the performance, the interference power at the relays determines $G_{c}$, which is better than that when the second hop is dominating and the interference at the destination is more severe on the behavior. 
\restylefloat{figure}

\begin{figure}[htb!]
\centering
\includegraphics[scale=0.34]{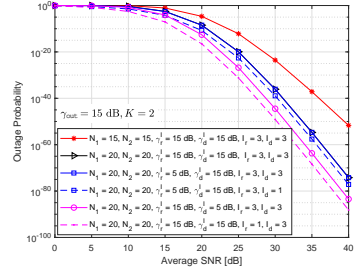}
\caption{$P_{\sf out}$ vs SNR for various values of $N_{1}$ and $N_{2}$ and various interference parameters.}\label{Pout_SNR_N1_N2_gIr_gId_Ir_Id}
\end{figure}

Fig. \ref{Pout_SNR_N1_N2_gIr_gId_Ir_Id} studies the case when $N_{1}=N_{2}$, where the interference power is determining which hop dominates the system performance. When $N_{1}=N_{2}=15$, $G_{d}=30$, whereas when $N_{1}=N_{2}=20$, $G_{d}=40$. Clearly, when the interference power is reduced at the relays, the enhancement on $G_{c}$ is unnoticeable, whereas when the interference power is reduced at the destination, $G_{c}$ increases and the system gain becomes noticeable. This behavior is expected as the interference at the destination is more impactful on the performance than the relays interference. We can also see that when the first hop is dominating the performance ($\bar{\gamma}^{I}_{\sf r}=15\ \mathrm{dB}, \bar{\gamma}^{I}_{\sf d}=5\ \mathrm{dB}$), decreasing the number of interferers at the relays from 3 to 1 has a noticeable impact on $G_{c}$. Same thing applies when the second hop is dominating the performance ($\bar{\gamma}^{I}_{\sf r}=5\ \mathrm{dB}, \bar{\gamma}^{I}_{\sf d}=15\ \mathrm{dB}$), where decreasing the number of interferers at the destination from 3 to 1 also noticeably enhances $G_{c}$.  
\restylefloat{figure}

%%%%%%%%%%%%%%%%%%%%%%%%%%%%%%%%%%%%%%%%%%%%%%%%%%%%%%%%%%%%%%%%%%%%%%%%%%%%%%%%%%%%%%
%%%%%%%%%%%%%%%%%%%%%%%%%%%%%% Section 5: Conclusion %%%%%%%%%%%%%%%%%%%%%%%%%%%%%%%%%
%%%%%%%%%%%%%%%%%%%%%%%%%%%%%%%%%%%%%%%%%%%%%%%%%%%%%%%%%%%%%%%%%%%%%%%%%%%%%%%%%%%%%%
\section{Conclusion}\label{C}
This letter considered the scenario of DF relay network with RIS-assisted source and relays with interference. It derived approximate closed-form and asymptotic expressions for the outage probability assuming Rayleigh fading channels. Results showed that the system can achieve a diversity order of $G_{d}=\mathrm{min}(N_{1},N_{2})K$. Findings illustrated that for the same $G_{d}$, utilizing one relay with multiple reflecting elements gives better performance than utilizing multiple relays with a single reflecting element. The interference at the destination is more severe on the performance than the interference at the relays. Therefore, under the same interference and for a given $K$, the case $N_{1}<N_{2}$ outperforms the case $N_{2}<N_{1}$ in terms of $G_{c}$.

%In this paper, the outage performance of a dual-hop
%$N^{\mathrm{th}}$-best DF relay system was evaluated in the
%presence of interference at the relays and destination. We derived
%exact closed-form expression for the outage probability with all
%system links assumed to be Rayleigh distributed. Furthermore, the
%outage performance of the considered system was studied at high
%SNR regime via deriving the asymptotic outage probability.
%Compared with Monte-Carlo simulations, a perfect fitting between
%all curves can be noticed. Results show that the diversity order
%linearly increases with the number of relays and linearly
%decreases with the order of the relay. Finally, the asymptotic
%results show that the system is still able to achieve full
%diversity gain in the presence of finite number of interferers
%with finite powers.
%%%%%%%%%%%%%%%%%%%%%%%%%%%%%%%%%%%%%%%%%%%%%%%%%%%%%%%%%%%%%%%%%%%%%%%%%%%%%%%%%%%%%%
%%%%%%%%%%%%%%%%%%%%%%%%%%%% Section 6: Acknowledgement %%%%%%%%%%%%%%%%%%%%%%%%%%%%%%
%%%%%%%%%%%%%%%%%%%%%%%%%%%%%%%%%%%%%%%%%%%%%%%%%%%%%%%%%%%%%%%%%%%%%%%%%%%%%%%%%%%%%%
%\section*{Acknowledgement}
%====================================================================
%%%%%%%%%%%%%%%%%%%%%%%%%%%%%%% Appendices %%%%%%%%%%%%%%%%%%%%%%%%%%%%%
%%%%%%%%%%%%%%%%%%%%%%%%%%%%%%%%%%%%%%%%%%%%%%%%%%%%%%%%%%%%%%%%%%%%%%%%%%%%%%%%%%%%%%
%%%%%%%%%%%%%%%%%%%%%%%%%%%%%%%%%% Appendix I %%%%%%%%%%%%%%%%%%%%%%%%%%%%%%%%%%%%%%%%
%%%%%%%%%%%%%%%%%%%%%%%%%%%%%%%%%%%%%%%%%%%%%%%%%%%%%%%%%%%%%%%%%%%%%%%%%%%%%%%%%%%%%%
\appendices
\section{Proof of Theorem \ref{theorem:1}}\label{appendix:1}
In this Appendix, we evaluate the first term in (\ref{Eq.9})
$\mathrm{P_{r}}\left[\gamma_{\sf d}<u|B_{L}\right]$. First, the e2e SINR
can be written as $\gamma_{\sf d} = Y/Z$. The CDF of
$\gamma_{\sf d}$ given a decoding set $B_{L}$ is given by
$\mathrm{P_{r}}\left[\gamma_{\sf
d}<u|B_{L}\right]=\int_{1}^{\infty}f_{Z}(z)\int_{0}^{uz}f_{Y}(y)dydz = \int_{1}^{\infty}f_{Z}(z)F_{Y}(uz)dz$. With RIS-assisted relays of $N_{2}$ reflecting elements, the CDF of the second hop SNR is given by \cite{Yang6}
\begin{align}\label{sese}
F_{\gamma_{\sf r,d}}(\gamma) = 1- e^{-\frac{\lambda_{\sf r,d} \gamma}{C}} \sum_{k = 0}^{N_{2} - 1} \frac{(\lambda_{\sf r,d} \gamma)^k}{C^k k!},
\end{align}
where $C = 1 + (N_{2} - 1)\Gamma^{2}\left(\frac{3}{2} \right)$. Upon utilizing the Binomial rule, the CDF of the best selected relay can be rewritten as
\begin{align*}
&F_{Y}(uz)= (F_{\gamma_{\sf r,d}}(uz))^L=\left(1- e^{-\frac{\lambda_{\sf r,d} uz}{C}} \sum_{k = 0}^{N_{2} - 1} \frac{(\lambda_{\sf r,d} uz)^k}{C^k k!}\right)^L\\
&= \sum_{k=0}^{L} \begin{pmatrix}
L \\ k \end{pmatrix} (-1)^{k} \exp\left(-\lambda_{\sf r,d}u z k\right) \sum_{j_{1}=0}^{N_{2}-1}...\sum_{j_{k}=0}^{N_{2}-1}u^{\sum_{n=1}^{k}j_{n}} \nonumber\\
&\times \frac{1}{(C/\lambda_{\sf r,d})^{\sum_{n=1}^{k}j_{n}}\prod_{n=1}^{k}j_{n}!} .
\end{align*}
With Rayleigh distributed interference, the interference-to-noise ratio (INR) CDFs at {\sf D} will be following exponential distribution as $F_{\gamma_{\sf d}^{I}}(\gamma)=1-e^{-\lambda_{\sf d}^{I}\gamma}$. Hence, the RV $Z$, which equal to $Z=X+1$, where $X$ is the interference at the destination has the following PDF \cite{Anas}
\begin{align}\label{Eq.13w}
f_{X}(x)= - \frac{\left(\lambda_{\sf
 d}^{I} \right)^{I_{d}}}{\left(I_{d} - 1 \right)!}x^{I_{d}-1}e^{-\lambda_{\sf d}^{I}x}.
\end{align}
Using transformation of RVs and then the binomial rule, the PDF of $Z$ can be obtained as
\begin{align}\label{Eq.13}
f_{Z}(z)= - \frac{\left(\lambda_{\sf
 d}^{I} \right)^{I_{d}}}{\left(I_{d} - 1 \right)!}\frac{\left(-1 \right)^{I_{d}}}{e^{-\lambda_{\sf d}^{I}}}\sum_{g=0}^{I_{d} - 1}\ \begin{pmatrix}
I_{d} - 1 \\ g
\end{pmatrix} \frac{(-1)^{g} z^{g}}{e^{\lambda_{\sf d}^{I}z}}.
\end{align}
Now, using the integral $\int_{1}^{\infty}f_{Z}(z)F_{Y}(uz)dz$ and with the help of \cite[Eq. (8.351.2)]{Grad.}, the first term in \eqref{Eq.9} $\mathrm{P_{r}}\left[\gamma_{\sf d}<u|B_{L}\right]$ can be obtained as in \eqref{Eq.27az}.

To obtain the second term in (\ref{Eq.9}) $\mathrm{P_{r}}\left[\gamma_{{\sf s},k}<u\right]$, the first hop SINR can be written as $Y'/Z'$, where $Y'$ is the first hop SNR with RIS-aided transmitter and $Z'=X'+1$, where $X'$ is the interference at the relays. Again, the CDFs of the interferers at the relays follow exponential distribution as $F_{\gamma_{k}^{I}}(\gamma)=1-e^{-\lambda_{k}^{I}\gamma}, k=1...K$. With RIS-assisted source of $N_{1}$ reflecting elements, the CDF of first hop SNR is given by \cite{Yang6}
\begin{align}
F_{Y'}(uz) = 1- e^{-\frac{\lambda_{{\sf s},k} u z}{B}} \sum_{i = 0}^{N_{1} - 1} \frac{(\lambda_{{\sf s},k} u z)^i}{B^i i!},
\end{align}
where $B = 1 + (N_{1} - 1)\Gamma^{2}\left(\frac{3}{2} \right)$.\\
The PDF of $Z'$ is similar to that of $Z$, but with replacing $\lambda^{I}_{\sf d}$ with $\lambda^{I}_{k}$ and $I_{d}$ with $I_{k}$. Again, upon using the integral $\int_{1}^{\infty}f_{Z'}(z)F_{Y'}(uz)dz$ and with the help of \cite[Eq. (8.351.2)]{Grad.}, the second term in \eqref{Eq.9} $\mathrm{P_{r}}\left[\gamma_{{\sf s},k}<u\right]$ can be obtained as in \eqref{Eq.27bz}.

%%%%%%%%%%%%%%%%%%%%%%%%%%%%%%%%%%%%%%%%%%%%%%%%%%%%%%%%%%%%%%%%%%%%%%%%%%%%%%%%%%%%%%
%%%%%%%%%%%%%%%%%%%%%%%%%%%%%%%%%%% Appendix II %%%%%%%%%%%%%%%%%%%%%%%%%%%%%%%%%%%%%%
%%%%%%%%%%%%%%%%%%%%%%%%%%%%%%%%%%%%%%%%%%%%%%%%%%%%%%%%%%%%%%%%%%%%%%%%%%%%%%%%%%%%%%
\section{Proof of Theorem \ref{Theorem:2}}\label{appendix:3}
To find the asymptotic outage probability, we first need to
obtain $\mathrm{P_{r}}\left[\gamma_{\sf d}<u|B_{L}\right]$. As
$\rho\rightarrow\infty$ and with constant values of $\rho_{I}$,
$I_{k}$, and $I_{d}$, the CDF of the second hop channels with opportunistic relaying can be approximated as
\begin{align}\label{C.1}
F_{\gamma_{\sf r,d}}(\gamma)\approx
\frac{\gamma^{N_{2}}}{(C/ \lambda_{\sf r,d})^{N_{2}} N_{2}!}.
\end{align}
Now, the CDF of best selected relay can be written as
\begin{align}\label{C.1w}
F_{Y}(u z)\approx
\frac{(u z)^{LN_{2}}}{(C/ \lambda_{\sf r,d})^{LN_{2}} (N_{2}!)^{L}}.
\end{align}
With no change in the PDF of $Z$ and upon following the same procedure as in Appendix \ref{appendix:1},
the term $\mathrm{P_{r}}\left[\gamma_{\sf d}<u|B_{L}\right]$ in
(\ref{Eq.9}) can be evaluated at high SNR values as in \eqref{Eq.28a}.

Now, the second term in \eqref{Eq.9}
$\mathrm{P_{r}}\left[B_{L}\right]$ can be obtained after
evaluating the CDF of $\gamma_{{\sf s},k}$, which can be
approximated at high SNR values as
\begin{align}\label{C.1}
F_{\gamma_{{\sf s},k}}(u z)\approx
\frac{(u z)^{N_{1}}}{(B/ \lambda_{{\sf s},k})^{N_{1}} N_{1}!}.
\end{align}
Again, with no change in the PDF of $Z'$ and upon following the same procedure as in Appendix \ref{appendix:1},
the term $\mathrm{P_{r}}\left[B_{L}\right]$ in
(\ref{Eq.9}) can be evaluated at high SNR values as in \eqref{Eq.28b}.

%====================================================================

%\newpage
%%%%%%%%%%%%%%%%%%%%%%%%%%%%%%%%%%%%%%%%%%%%%%%
%%%%%%%%%%%%% Balancing References %%%%%%%%%%%%
\balance
%%%%%%%%%%%%%%%%%%%%%%%%%%%%%%%%%%%%%%%%%%%%%%%
%%%%%%%%%%%%%%%%%%%%%%%%%%%%%%%%%%%%%%%%%%%%%%%

%%%%%%%%%%%%%%%%%%%%%%%%%%%%%%%%%%%%%%%%%%%%%%%%%%%%%%%%%
%%%%%%%%%%%%%%%%%%%%%%% Figures %%%%%%%%%%%%%%%%%%%%%%%%%
%%%%%%%%%%%%%%%%%%%%%%%%%%%%%%%%%%%%%%%%%%%%%%%%%%%%%%%%%

%\begin{figure}
%\centering
%\includegraphics[width=3in]{figuree.eps}
%% where an .eps filename suffix will be assumed under latex,
%% and a .pdf suffix will be assumed for pdflatex; or what has been declared
%% via \DeclareGraphicsExtensions.
%\caption{Symbol error probability performance of
%interference-limited destination fixed-gain AF systems with
%various interfering fading channels at $\rho$ = 17
%dB.}\label{figure2}
%\end{figure}

%====================================================================

\end{document}